\documentclass[12pt,preprint]{aastex}

\usepackage{natbib}
\newcommand{\Fermi}{{\textit{Fermi}}}

\slugcomment{Accepted for publication in ApJ Letters}
\begin{document}

\title{Fermi Observations of GRB~090902B: A Distinct Spectral
       Component in the Prompt and Delayed Emission}

\author{
A.~A.~Abdo\altaffilmark{1,2}, 
M.~Ackermann\altaffilmark{3}, 
M.~Ajello\altaffilmark{3}, 
K.~Asano\altaffilmark{4,5}, 
W.~B.~Atwood\altaffilmark{6}, 
M.~Axelsson\altaffilmark{7,8}, 
L.~Baldini\altaffilmark{9}, 
J.~Ballet\altaffilmark{10}, 
G.~Barbiellini\altaffilmark{11,12}, 
M.~G.~Baring\altaffilmark{13}
% ...
}
\altaffiltext{1}{Space Science Division, Naval Research Laboratory, Washington, DC 20375, USA}
\altaffiltext{2}{National Research Council Research Associate, National Academy of Sciences, Washington, DC 20001, USA}
\altaffiltext{3}{W. W. Hansen Experimental Physics Laboratory, Kavli Institute for Particle Astrophysics and Cosmology, Department of Physics and SLAC National Accelerator Laboratory, Stanford University, Stanford, CA 94305, USA}
\altaffiltext{4}{Department of Physics, Tokyo Institute of Technology, Meguro City, Tokyo 152-8551, Japan}
\altaffiltext{5}{Interactive Research Center of Science, Tokyo Institute of Technology, Meguro City, Tokyo 152-8551, Japan}
\altaffiltext{6}{Santa Cruz Institute for Particle Physics, Department of Physics and Department of Astronomy and Astrophysics, University of California at Santa Cruz, Santa Cruz, CA 95064, USA}
\altaffiltext{7}{Department of Astronomy, Stockholm University, SE-106 91 Stockholm, Sweden}
\altaffiltext{8}{The Oskar Klein Centre for Cosmoparticle Physics, AlbaNova, SE-106 91 Stockholm, Sweden}
\altaffiltext{9}{Istituto Nazionale di Fisica Nucleare, Sezione di Pisa, I-56127 Pisa, Italy}
\altaffiltext{10}{Laboratoire AIM, CEA-IRFU/CNRS/Universit\'e Paris Diderot, Service d'Astrophysique, CEA Saclay, 91191 Gif sur Yvette, France}
\altaffiltext{11}{Istituto Nazionale di Fisica Nucleare, Sezione di Trieste, I-34127 Trieste, Italy}
\altaffiltext{12}{Dipartimento di Fisica, Universit\`a di Trieste, I-34127 Trieste, Italy}
\altaffiltext{13}{Rice University, Department of Physics and Astronomy, MS-108, P. O. Box 1892, Houston, TX 77251, USA}
% ...


\begin{abstract}
We report on the observation of the bright, long gamma-ray burst,
GRB~090902B, by the Gamma-ray Burst Monitor (GBM) and Large Area
Telescope (LAT) instruments on-board the \Fermi\ observatory. This was
one of the brightest GRBs to have been observed by the LAT, which
detected several hundred photons during the prompt phase.  With a
redshift of $z = 1.822$, this burst is among the most luminous
detected by \Fermi.  Time-resolved spectral analysis reveals a
significant power-law component in the LAT data that is distinct from
the usual Band model emission that is seen in the sub-MeV energy
range. This power-law component appears to extrapolate from the GeV
range to the lowest energies and is more intense than the Band
component both below $\sim$\,50\,keV and above 100\,MeV.  The Band
component undergoes substantial spectral evolution over the entire
course of the burst, while the photon index of the power-law component
remains constant for most of the prompt phase, then hardens
significantly towards the end.  After the prompt phase, power-law
emission persists in the LAT data as late as 1 ks post-trigger, with
its flux declining as $t^{-1.5}$.  The LAT detected a photon with the
highest energy so far measured from a GRB, $33.4_{-3.5}^{+2.7}$ GeV.
This event arrived 82 seconds after the GBM trigger and
$\sim$50\,seconds after the prompt phase emission had ended in the GBM
band. We discuss the implications of these results for models of GRB
emission and for constraints on models of the Extragalactic Background
Light.
\end{abstract}


\section{Introduction}

The \Fermi\ Gamma-ray Space Telescope hosts two instruments, the Large
Area Telescope \citep{2009ApJ...697.1071A} and the Gamma-ray Burst
Monitor \citep{Meegan_GBM}, which together are capable of measuring
the spectral parameters of gamma-ray bursts (GRBs) across seven
decades in energy.  Since the start of GBM and LAT science operations
in early August 2008, emission at energies $>$100\,MeV has been
detected from ten GRBs.  These detections were made possible by the
LAT's greater sensitivity and shorter deadtime (26\,$\mu$s) compared
to previous instruments.  Prior to \Fermi, high-energy
gamma-rays from GRBs with energies up to 18 GeV were observed by the
EGRET instrument on-board the \textit{Compton Gamma-ray Observatory}.
The EGRET observations suggested three types of high-energy emission:
an extrapolation of the low energy spectra to the $>$100 MeV band
\citep[e.g.,][]{1998AIPC..428..349D}, an additional spectral component
during the prompt emission
\citep{2003Natur.424..749G,2008ApJ...677.1168K} and in the case of GRB
940217, a GeV afterglow which was detectable for 90 minutes after the
trigger \citep{1994Natur.372..652H}.  The redshifts of these events
were not determined.  Recently, \citet{2008A&A...491L..25G} reported
that GRB~080514B which triggered AGILE at lower energies, was detected
by the GRID instrument up to 300 MeV. A photometric redshift of
$z=1.8^{+0.4}_{-0.3}$ was reported for this event
\citep{2008A&A...491L..29R}.

In the \Fermi\ era, due to the advanced localization capabilities of
the LAT and the rapid follow-up by the \textit{Swift} narrow field
instruments \citep{2004ApJ...611.1005G} and the ground-based follow-up
community, redshifts for five of the ten LAT bursts have been
measured.  These include GRB~080916C with $z = 4.35 \pm 0.15$
\citep{2009A&A...498...89G}, a long burst that has the highest
inferred isotropic energy, $E_{\rm iso} \approx 8.8 \times
10^{54}$\,ergs (10 keV--10 GeV) \citep{2009Sci...323.1688A}, and
GRB~090510 with $z = 0.903 \pm 0.003$ \citep{2009_arne}, the second
short burst seen by the LAT and the first short burst to show
definitively an additional hard power-law component in the GeV band
during the prompt phase \citep{Nature..GRB090510}.  

GRB~090902B is a long, fairly intense burst with a redshift of $z =
1.822$ \citep{Redshift_090902B} and fluence of $(4.36 \pm 0.06) \times
10^{-4} $\,erg\,cm$^{-2}$ (10 keV--10 GeV) over the first 25 seconds
of the prompt emission.  These data give an isotropic energy $E_{\rm
  iso} = (3.63\pm 0.05) \times 10^{54}$\,ergs, comparable to that of
GRB~080916C.  Similar to GRB~090510, GRB~090902B has a significant
additional, hard power-law component that appears during the prompt
phase.  Furthermore, a spectral feature at energies $\la 50$\,keV is
evident in the GBM spectrum of GRB~090902B that is consistent with an
extrapolation of the $>$\,100\,MeV power-law emission down to those
energies.  In previous analyses, \citet{Preece:1996} reported evidence
for an additional low-energy spectral component below 20\,keV for
$\sim 15$\% of BATSE bursts.

We report on the observations and analysis of gamma-ray emission from
GRB~090902B measured by the GBM and LAT instruments.  In section 2, we
present details of the detections by both instruments and summarize
the follow-up observations. In section 3, we show the light curves of
the prompt emission as seen by the various detectors and describe the
extended emission found in the LAT data out to 1 ks after the trigger.
In section 4, we present the time-resolved spectral analysis of the
burst emission during the prompt phase. Finally, in section 5, we
discuss the physical interpretation of the GBM and LAT data, focusing
on the implications of the power-law component for models
of GRB physics.

\section{Observations}

On 2009 September 2 at 11:05:08.31 UT, the \Fermi\ Gamma-ray Burst
Monitor triggered on and localized the bright burst GRB 090202B
\citep[trigger 273582310 / 090902462,][]{Betta_090902B}.  The burst
was within the LAT field of view initially at an angle of 51$^\circ$
from the boresight.  This event was sufficiently bright in the GBM
that an Autonomous Repoint Request was made, and the spacecraft began
slewing within 10 seconds towards the burst.  After $\sim$200 seconds,
it had pointed the LAT boresight to within a few degrees of the final
burst localization. It maintained that pointing until $\sim$1 ks
post-trigger, when the Earth's limb began to enter the LAT
field-of-view (FOV).  This burst was detected up to $\sim$5 MeV by
GBM, and emission was significantly detected by the LAT, with 39
photons above 1 GeV.  The highest energy photon
had $E = 33.4_{-3.5}^{+2.7}$ GeV and arrived 82 seconds after the GBM
trigger; and the initial analyses detected photons as late as 300
seconds after the trigger \citep{LAT_090902B}.

From the LAT data, the burst was localized to R.A.(J2000), Dec(J2000)
= 265.00, 27.33 with a statistical uncertainty of 0.04$^\circ$ (+
$<$0.1$^\circ$ systematic), enabling Target of
Opportunity observations to begin $\sim$\,12.5 hours after the trigger
with the narrow field instruments on \textit{Swift}.  A candidate
X-ray afterglow within the LAT error circle was detected by the X-Ray
Telescope \citep[XRT,][]{Swift_XRT_090902B}. This source was confirmed
to be fading \citep{Swift_XRT2_090902B}, and UVOT observations
revealed the optical afterglow
\citep{Swift_UVOT2_090902B}.  The earliest ground-based optical
observations were obtained by ROTSE-IIIa $\sim$1.4 hours post trigger
\citep{ROTSE_090902B}. Other detections were reported in the optical
\citep{Perley_090902B}, in the near infrared by GROND
\citep{GROND_090902B} and in the radio
\citep{WSRT_090902B,VLA_090902B}.  The location of the fading source
detected by GROND was R.A.(J2000), Dec(J2000) = $17^{\rm h}39^{\rm
m}45\fs41$, +27\arcdeg 19\arcmin 27.1\arcsec, 3.3 arcminutes from the
LAT location \citep{GROND_090902B}.  The afterglow redshift of $z =
1.822$ was measured by \citet{Redshift_090902B} using the GMOS
spectrograph mounted on the Gemini-North telescope.

\section{Light Curves}

\begin{figure}
  \centering
  \includegraphics[scale=0.7]{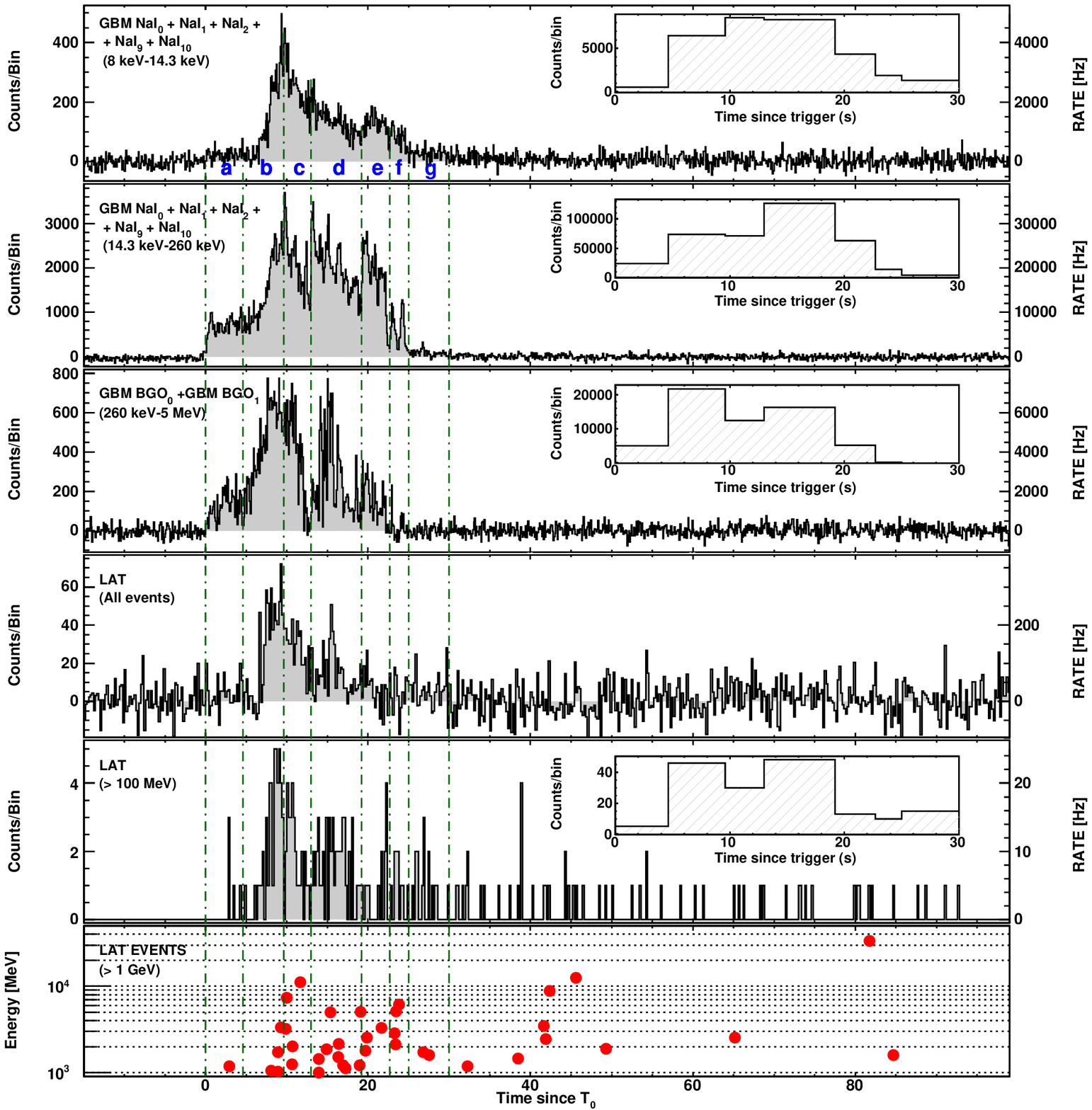}
  \caption{GBM and LAT light curves for the gamma-ray emission of GRB
  090902B.  The data from the GBM NaI detectors were divided into soft
  (8--14.3 keV) and hard (14.3--260 keV) bands in order to reveal any
  obvious similarities between the light curve at the lowest energies
  and that of the LAT data.  The fourth panel shows all LAT events
  that pass the on-board gamma filter, while the fifth and sixth
  panels show data for the ``transient'' class event selection for
  energies $>$\,100\,MeV and $>$\,1\,GeV, respectively.  The vertical
  lines indicate the boundaries of the intervals used for the
  time-resolved spectral analysis.  Those time boundaries are at $T_0
  + (0, 4.6, 9.6, 13.0, 19.2, 22.7, 25.0, 30.0)$ seconds.  The insets
  show the counts for the corresponding dataset binned using these
  intervals in order to illustrate the relative numbers of counts
  considered in each spectral fit.}
  \label{Fig:prompt light curves}
\end{figure}

In Figure~\ref{Fig:prompt light curves}, we show the GBM and LAT light
curves in several energy bands.  The top three panels show data from
the most brightly illuminated NaI and BGO detectors of the GBM, and
the bottom three panels show the LAT data with various event
selections.  In the bottom panel, the measured photon energies are
plotted as a function of time, including the highest energy event ($E
= 33.4$\,GeV) that arrived 82 seconds after the GBM trigger time,
$T_0$.  From the GBM light curves, we see that at energies
$\la$\,1\,MeV the prompt phase ends approximately 25 seconds after
$T_0$.  Detailed analysis of the GBM data for energies 50--300 keV
yields a formal T90 duration\footnote{The T90 duration is the time
  over which the central 90\% of the counts from the burst have been
  accumulated.} of 21.9 seconds starting at $T_0 + 2.2$\,s. By
contrast, the LAT emission $>$100\,MeV clearly continues well after
this time range.

\begin{figure}
  \centering
  \plotone{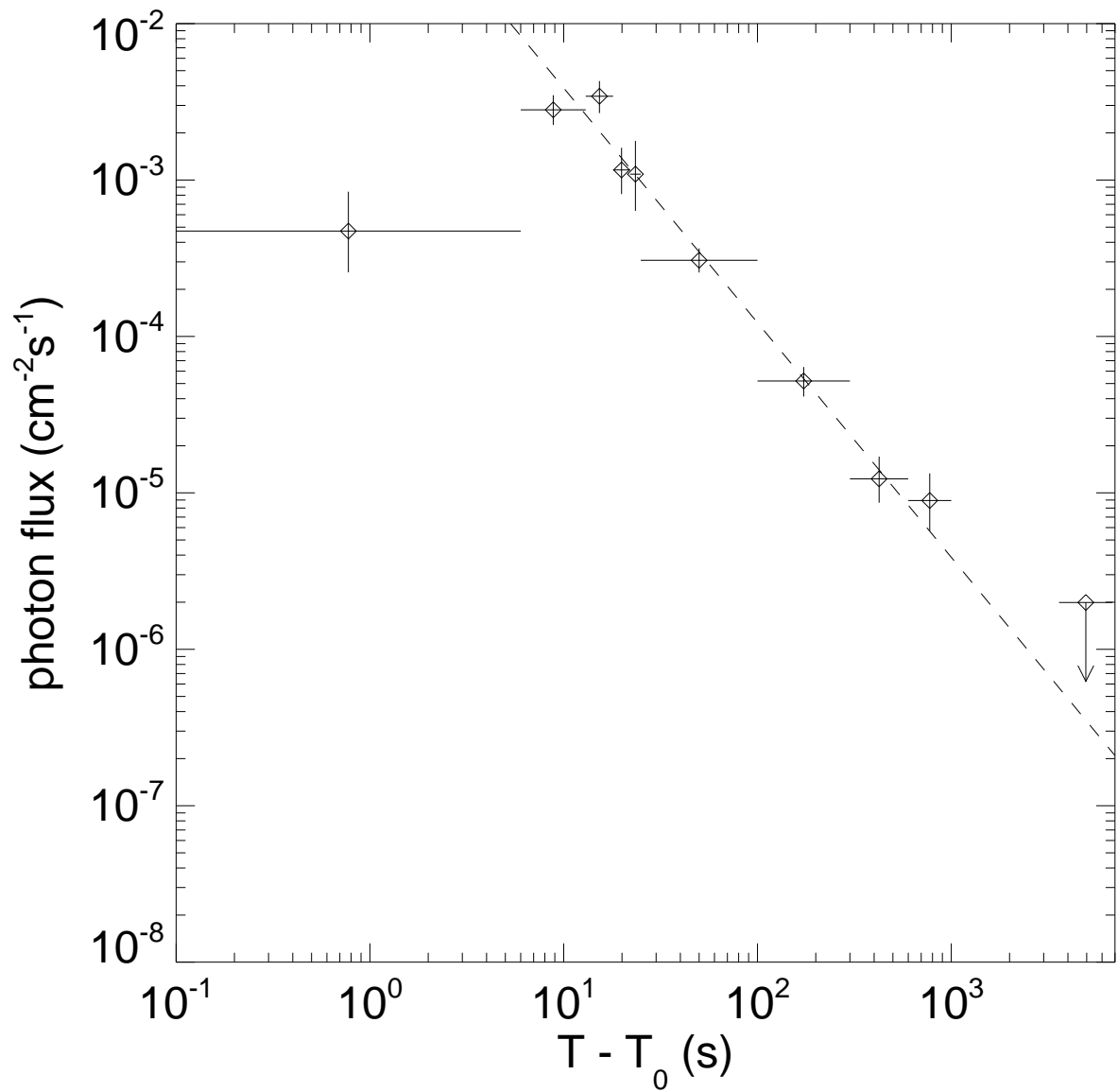}
  \caption{Light curve of GRB 090902B for energies 0.1--300 GeV from
    unbinned likelihood fits to the LAT data. After the prompt phase,
    extended or afterglow emission consistent with a temporal profile
    $\propto t^{-1.5}$ (dashed line) lasts until $\sim T_0 + 1000$\,s.
    The upper limit at times $> T_0 + 3600$\,s was derived from the data
    collected after the source emerged from occultation by the Earth. }
  \label{Fig:GeV afterglow}
\end{figure}

On time scales longer than the prompt phase, the LAT detects emission
from GRB 090902B as late as 1 ks after the GBM trigger.  The spectrum
of this emission is consistent with a power-law with photon index
$\Gamma = -2.1 \pm 0.1$, and its flux ($>$100\,MeV) declines as 
$t^{-1.5 \pm 0.1}$ over the interval ($T_0 + 25, T_0 + 1000$\,s).  As we note
above, the LAT observations are interrupted by entry of the Earth's
limb into the FOV, but analysis of data after $T_0 + 3600$\,s, when
the source location is again unocculted, shows that any later emission
lies below the LAT sensitivity (Figure~\ref{Fig:GeV afterglow}).  The
upper limit we obtain for data after $T_0 + 3600$ is consistent with
an extrapolation of the $t^{-1.5}$ decay.  Similar late time emission
for energies $>$100\,MeV that extends well beyond the prompt phase has
been seen for five earlier bursts by \Fermi: GRB~080916C
\citep{2009Sci...323.1688A}; GRB~090323 \citep{2009GCN..9021....1O};
GRB~090328 \citep{2009GCN..9077....1C}; GRB~090510, independently seen
by AGILE \citep{2009arXiv0908.1908G} and by
\Fermi\ \citep{2009arXiv0909.0016G}; and GRB~090626
\citep{2009GCN..9584....1P}.

\section{Time-resolved Spectral Analysis}

Spectral analysis was performed using the data from both the GBM and
the LAT. These analyses include data from the NaI detectors 0,1,2,9,10
and BGO detectors, and LAT ``transient'' class data, with front- and
back-converting events considered separately. The NaI data are fit
from 8 keV to 1 MeV and the BGO from 250 keV to 40 MeV using the Time
Tagged Event (TTE) data, which are high time resolution data that
allow us to define the time intervals based on the structure of the
GBM and LAT light curves.  The LAT data are fit from 100 MeV to 200
GeV.  An effective area correction of 0.9 has been fit to the BGO data
to match the model normalizations given by the NaI data; this
correction is consistent with the uncertainties in the GBM detector
responses.  The fits were performed with the spectral analysis
software package RMFIT (version 3.1). For further details on the data
extraction and spectral analysis procedures see
\citet{Fermi..GRB080825C} and \citet{Fermi..GRB081024B}.

\begin{figure}
  \centering
  \includegraphics[scale=0.5,bb=-40 10 550 690,clip]{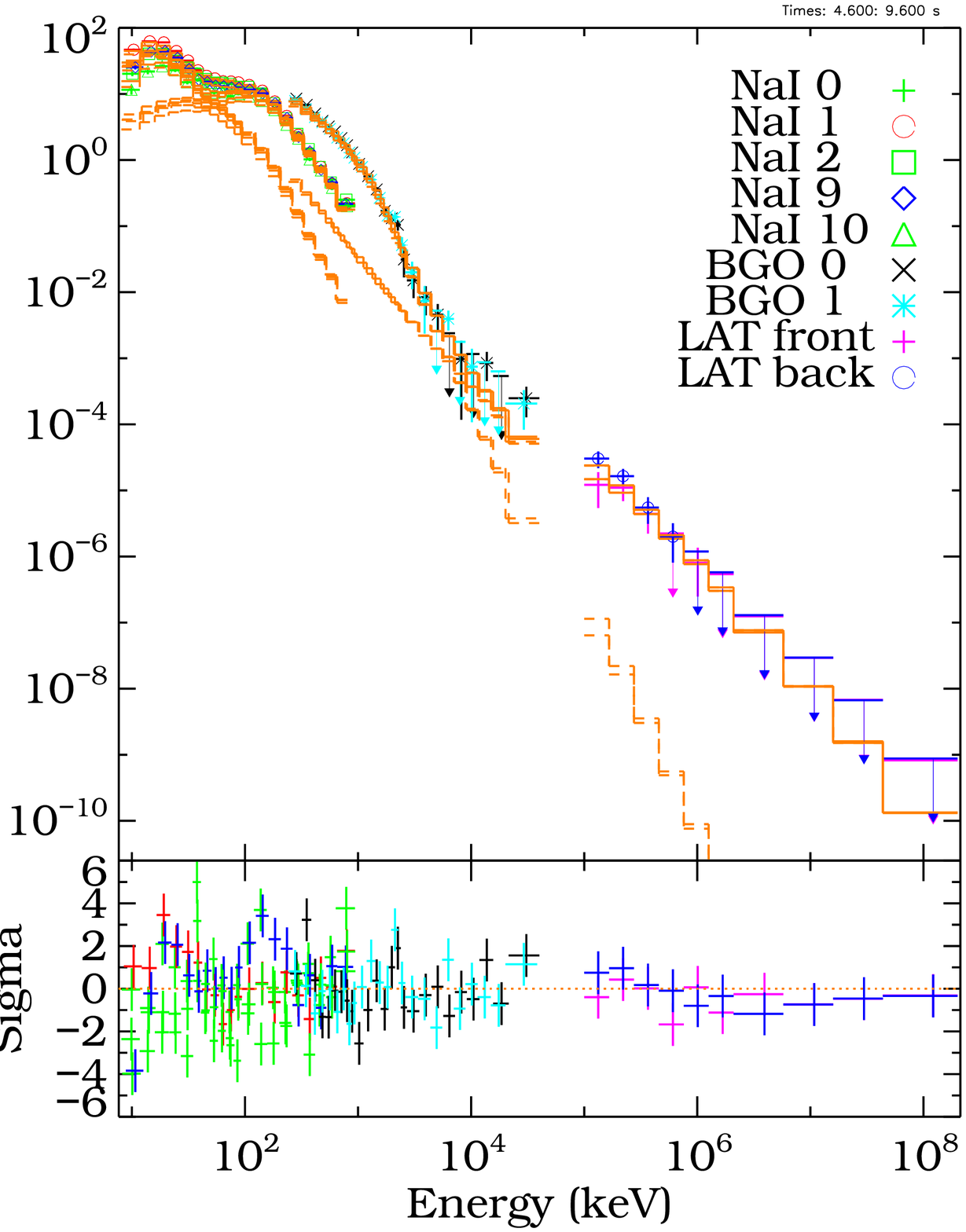}\\
  \includegraphics[scale=0.5,clip]{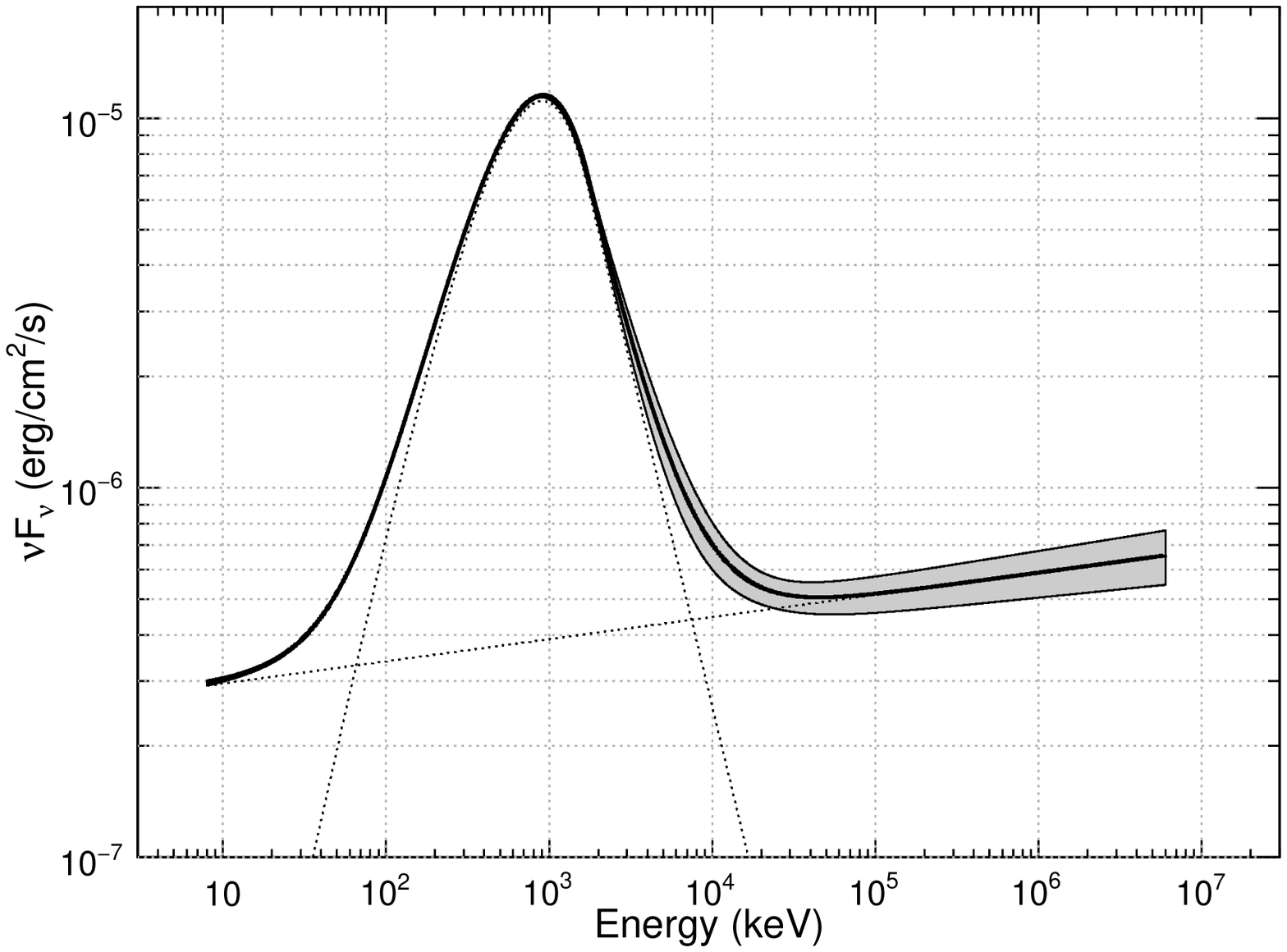}
  \caption{Joint fit of GBM and LAT data to interval \textbf{b}, ($T_0
    + 4.6, T_0 + 9.6$\,s).  Top: Counts spectrum; separate model
    components are plotted, Band (dashed), power-law (solid).  Bottom:
    Unfolded $\nu F_\nu$ spectrum.  The extension of the $> 100$\,MeV
    power-law component to the lowest energies ($< 50$\,keV) is
    shown.}
  \label{Fig:prompt spectrum}
\end{figure}

The time-integrated spectrum of GRB~090902B is best modeled by a Band
function \citep{Band:93} and a power-law component
(Table~\ref{Tab:spectral params}).  The power-law component
significantly improves the fit between 8\,keV and 200\,GeV both in the
time-integrated spectrum and in the individual time intervals where
there are sufficient statistics.  It is also required when considering
only the GBM data (8\,keV--40\,MeV) for the time-integrated spectrum,
as its inclusion causes an improvement of $\approx$\,2000 in the CSTAT
statistic over the Band function alone.  When data below $\sim$\,50
keV are excluded, a power-law component can be neglected in the
GBM-only fits.  We conclude that this power-law component
contributes a significant part of the emission both at low
($<$\,50\,keV) and high ($>$\,100\,MeV) energies.
Figure~\ref{Fig:prompt spectrum} shows the counts and unfolded $\nu
F_\nu$ spectra for a Band function with a power-law component fit to
the data for interval \textbf{b} (when the low energy excess is most
significant) using the parameters given in Table~\ref{Tab:spectral
  params}.

Spectral evolution is apparent in the Band function component from the
changing $E_{\rm peak}$ values throughout the burst, while $\beta$
remains soft until interval \textbf{e} when it hardens significantly.
$\beta$ is similarly hard in interval \textbf{f}, after which the Band
function component is no longer detected.  The hardening of $\beta$ is
accompanied by an apparent hardening of the power-law index, $\Gamma$,
which until interval \textbf{e} does not exhibit much variation.
However, this is not definitive since the flux is too low to constrain
$\Gamma$ in intervals \textbf{e} and \textbf{f} separately.  A
spectral fit of the sum of these two intervals confirms the presence
of both a harder $\beta$ and a harder $\Gamma$, with a clear
statistical preference for the inclusion of the power-law component.
An equally good fit is obtained in the combined
\textbf{e}~+~\textbf{f} interval if this power-law has an exponential
cut-off at high energies, with the preferred cut-off energy lying
above 2\,GeV.  Finally, we note that in interval \textbf{b}, a
marginally better fit is achieved using a model with the additional
power-law component having an exponential cut-off at high
energies. The improvement is at the $\sim 3\sigma$ level and indicates
weak evidence for a cutoff in the second component, placing a lower
limit on the cutoff energy in this interval of about 1\,GeV.

\section{Discussion and Interpretation}

The \Fermi\ data for GRB~090902B show for the first time clear
evidence of excess emission both at low energies ($\la 50\;$keV) and
at high energies ($>$100\,MeV), while the Band function alone fits
data at intermediate energies adequately.  These excesses are well-fit
by a single power-law component suggesting a common origin.  This
power-law component accounts for $\approx 24\%$ of the total fluence
in the 10~keV--10~GeV range, and its photon index is hard, with a
value $\sim -1.9$ throughout most of the prompt phase.
Such a hard component producing
the observed excess at low energies is difficult to explain in the
context of leptonic models by the usual synchrotron self-Compton (SSC)
mechanisms. 

In the simplest versions of these models, the peak of the SSC emission
is expected to have a much higher energy than the synchrotron peak at
MeV energies, and the SSC component has a soft tail that is well below
the synchrotron flux at lower energies and so would not produce excess
emission below $\sim 50$~keV.  Hadronic models, either in the form of
proton synchrotron radiation \citep{2009arXiv0908.0513R} or
photohadronic interactions \citep{2009arXiv0909.0306A}, can produce a
hard component with a similar low energy excess via direct and cascade
radiation (e.g., synchrotron emission by secondary pairs at low
energies).  However, the total energy release in hadronic models would
exceed the observed gamma-ray energy of $E_{\rm iso} = 3.63\times
10^{54}$\,ergs significantly and may pose a challenge for the total
energy budget.  Collimation into a narrow jet may alleviate the energy
requirements, since the actual energy release from GRB~090902B can be
smaller by a jet beaming factor $> 1/\Gamma_0^2$ from the apparent
isotropic value, where $\Gamma_0$ is the bulk Lorentz factor of the
fireball.

From the observation of a $11.16^{+1.48}_{-0.58}$~GeV photon in
interval \textbf{c}, the highest energy during the prompt phase and
thus the most constraining, we derive a minimum value of the bulk
Lorentz factor $\Gamma_{\rm min} \approx 1000$ using the flux
variability time scale of $t_v \approx 53$\,ms found in the BGO data.
This limit follows from the constraint that the opacity for $e^\pm$
pair production with target photons fitted by the Band+PL model in
interval \textbf{c} is less than unity for the 11.16\,GeV photon~\citep[see,
e.g.,][]{1993A&AS...97...59F,1997ApJ...491..663B,2001ApJ...555..540L}. This
high $\Gamma_{\rm min}$ value is of the same order as the values
derived for GRB~080916C~\citep{2009Sci...323.1688A} and
GRB~090510~\citep{Nature..GRB090510}, both of which have been detected
at $> 10\;$GeV with the LAT.

The delayed onset of the $\gtrsim$100 MeV emission from the GBM
trigger has been modeled for GRB~080916C as arising from proton
synchrotron radiation in the prompt phase~\citep{2009arXiv0908.0513R}
and for GRB~090510 as arising from electron synchrotron radiation in
the early afterglow phase
\citep{2009arXiv0905.2417K,2009arXiv0909.0016G}.  In order to produce
the peak of the LAT emission at $\sim T_0+9$\,s in the early afterglow
scenario for GRB~090902B from deceleration of the GRB fireball, a
value of $\Gamma_0 \approx 1000$ is required.  This is similar to
$\Gamma_{\rm min}$ that we calculate, but the observed large amplitude
variability on short time scales ($\approx 90$~ms) in the LAT data,
which is usually attributed to prompt emission, argues against such
models.  Also, the appearance of the power-law component extending
down to $\approx 8$~keV within only a few seconds of the GRB trigger
disfavors an afterglow interpretation. The proton synchrotron model,
on the other hand, requires a rather large total energy budget, as
mentioned previously.

Yet another interpretation of the observed excess in the high and low
energies may be provided by two non-thermal power-law components along
with a thermal component from the jet
photosphere~\citep{2000ApJ...530..292M,2004ApJ...614..827R}.  The
thermal component, broadened by temperature variations, then accounts
for the $\gtrsim 100\;$keV--few MeV emission with $\Gamma_0 \approx
930$~\citep{2007ApJ...664L...1P}, although fits of such a model to our
data do not improve over the Band+PL model.  Furthermore, it is
difficult for the photospheric model to explain the delayed onset of
the $\ga$\,100\,MeV emission.

The detection of the $33.4\;$GeV photon, $82\;$seconds after the GRB
trigger and well after the soft gamma-ray emission subsided, may help
constrain the origin of the late-time decay of the power-law
component, which goes as $t^{-1.5}$.  A synchrotron origin of the 33.4
GeV photon would be difficult since it would require significant
energy gain by electrons over a gyroradius and a bulk Lorentz factor
$>$\,1500.  In the case of diffusive shock-acceleration, the energy
losses in the upstream region of the shock may dominate \citep[see,
e.g.,][]{2006ApJ...651..328L} and prevent acceleration of electrons to
an energy high enough to radiate a 33.4 GeV photon.  An interpretation
by afterglow SSC emission is still possible, however.

The constraints on the quantum gravity mass scale from GRB~090902B
using the time-of-flight test \citep{Amelino-Camelia:98} are much
weaker than those from GRB~090510 \citep{Nature..GRB090510} due to the
larger interval, 82 seconds, between $T_0$ and the arrival time of the
33.4\,GeV photon. However, the moderately high redshift ($z = 1.822$)
of GRB~090902B allows us to use this photon to probe and constrain
models of the Extragalactic Background Light \citep[EBL;
][]{2004A&A...413..807K,2003A&A...407..791M,2006ApJ...648..774S,2008A&A...487..837F,2009arXiv0905.1144G,2009arXiv0905.1115F}.
The 33.4 GeV photon would not be absorbed by the EBL in any models
except for the ``fast evolution'' and the ``baseline'' models by
\citet{2006ApJ...648..774S}, which give optical depths of
$\tau_{\gamma\gamma} = 7.7$ and 5.8, respectively.  We have performed
spectral fits of the LAT data with and without the predicted EBL
absorption from Stecker's models assuming a simple power-law as the
intrinsic emission model. Based on Monte-Carlo simulations, we found
that Stecker's fast evolution and baseline models are disfavored at a
$> 3\sigma$ level.

In summary, GRB~090902B is one of the brightest bursts detected by the
GBM and LAT instruments on \Fermi.  It clearly shows excess emission
at high and low energies during the prompt phase, requiring a hard
power-law component in addition to the usual Band function in order to
fit the data.  The origin of this component is not understood, and its
presence in this burst poses genuine challenges for the
theoretical models.  Like the other two bright \Fermi\ bursts detected
by the LAT, GRB 080916C and GRB 090510, GRB 090902B appears to possess
a very high Lorentz factor for the bulk outflow, $\Gamma \approx
1000$, and has some suggestion of a delayed onset of the emission
above $\sim$\,100 MeV.  Finally, the 33.4\,GeV photon, the highest
energy yet detected from a GRB, and the $z = 1.822$ redshift of this
burst have allowed us to place significant constraints on some models
of the Extragalactic Background Light.

\acknowledgments

The \Fermi\ LAT Collaboration acknowledges support from a number of
agencies and institutes for both development and the operation of the
LAT as well as scientific data analysis. These include NASA and DOE in
the United States, CEA/Irfu and IN2P3/CNRS in France, ASI and INFN in
Italy, MEXT, KEK, and JAXA in Japan, and the K.~A.~Wallenberg
Foundation, the Swedish Research Council and the National Space Board
in Sweden. Additional support from INAF in Italy and CNES in France
for science analysis during the operations phase is also gratefully
acknowledged.


\clearpage

\begin{deluxetable}{cccccccccccc}
\tabletypesize{\scriptsize}
\rotate
\tablecaption{Band function + power-law fit parameters for the  time-resolved spectral fits.}
\tablewidth{0pt}
\tablehead{
\colhead{Interval} & 
\colhead{Time Range (s) } & 
\colhead{E$_{\rm peak}$ (kev)} & 
\colhead{$\alpha$} &  
\colhead{$\beta$} & 
\colhead{$\Gamma$} &  
\colhead{CSTAT/DOF} & 
\colhead{$\Delta$CSTAT} & \colhead{Energy fluence}\\
 &  & &  &  & & & & \colhead{(erg\,cm$^{-2}$, 8\,keV--30\,GeV)}
 }
\startdata
\bf $\cdots$ &         0.0--30.0       &         726 ($\pm$8)&       -0.61 ($\pm$0.01)  & -3.8 ($^{+0.2}_{-0.3}$)        & -1.93($^{+0.01}_{-0.01}$)       & 2562/963       & 2005 &   (4.59 $\pm$ 0.05)$\times$10$^{-4}$  \\
\bf a. &         0.0--4.6       &         526 ($\pm$12)&                  -0.09 ($\pm$0.04)      & -3.7 ($^{+0.3}_{-0.6}$)        & -1.87($^{+0.04}_{-0.05}$)      & 901/963       & 43 &       (3.72 $\pm$ 0.13)$\times$10$^{-5}$  \\
\bf b.  & 4.6-- 9.6 &         908 ($^{+15}_{-14}$)   &  0.07 ($\pm$0.03) &          -3.9 ($^{+0.2}_{-0.3}$)     & -1.94 ($\pm$0.02)   &  1250/963 & 3165 &   (1.44 $\pm$ 0.03)$\times10^{-4}$  \\ 
\bf c. &  9.6--13.0 &  821 ($\pm$16) &       -0.26 ($\pm$0.03) &    -5.0($^{+0.8}_{-\infty}$)   &  -1.98 ($\pm$0.02) &        1310/963 & 2109 & (9.42 $\pm$ 0.24)$\times10^{-5}$ \\
\bf d.  &  13.0--19.2 &  529 ($\pm$9)        &  -0.65 ($\pm${-0.02}) &       -3.2 ($^{+0.1}_{-0.2}$)     &  -1.86 ($\pm$ 0.02) &        1418/963       & 199 &       (1.29 $\pm$0.03)$\times10^{-4}$ \\
\bf e.  & 19.2--22.7     &  317 ($\pm$8) &         -0.78 ($\pm${-0.02})  &  -2.4 ($\pm$0.1)   &   $\cdots$ &     1117/965       & $\cdots$ &  (4.8 $\pm$ 0.2) $\times10^{-5}$ \\
\bf f.  &          22.7--25.0 &         236 ($^{+25}_{-33}$) &  -1.30 ($^{+0.04}_{-0.03}$) &   -2.2 ($\pm$0.1)&      $\cdots$     &  1077/965 & $\cdots$ &   (1.0 $\pm$ 0.1)$\times 10^{-5}$ \\
\bf e.+f. & 19.2--25.0 & 327 ($\pm{8}$) & 	-0.91 ($\pm${0.02}) & -2.6 ($\pm{0.1}$) & -1.59 ($\pm{0.20}$) & 1219/963 & 16 & (6.1 $\pm$0.4)$\times 10^{-5}$ \\
\bf g.  &  25.0--30.0   & $\cdots$ & $\cdots$ & $\cdots$ &   -1.93 ($^{+0.25}_{-0.26}$)  &1209/967       &   $\cdots$ &     (6.8 $\pm$ 0.8)$\times 10^{-6}$ \\
\enddata

\newcommand{\pfrac}[2]{\left(\frac{#1}{#2}\right)}
\newcommand{\Epeak}{{E_{\rm peak}}}

\tablecomments{The time range values are relative to the trigger time
  $T_0$. The column $\Delta$CSTAT gives the change in CSTAT when fitting with only the 
  Band function versus Band+power-law. 
The Band function is given by 
\begin{eqnarray}
n(E) &=& A \pfrac{E}{100\,{\rm keV}}^\alpha 
           \exp\left(-\frac{E(2+\alpha)}{\Epeak}\right), \qquad E < E_c, 
           \nonumber\\
     &=& A \pfrac{(\alpha-\beta)\Epeak}{100\,{\rm keV}(2+\alpha)}^{\alpha-\beta}
           \exp(\beta - \alpha)\left(-\frac{E}{100\,{\rm keV}}\right)^\beta,
           \qquad E \ge E_c,
\label{eq:Band}
\end{eqnarray}
where $E_c = (\alpha - \beta)\Epeak/(2 + \alpha)$ \citep{Band:93}.  The power-law function is given by
\begin{equation}
n(E) = A \pfrac{E}{100\,{\rm keV}}^{\Gamma}.
\end{equation}}  
\label{Tab:spectral params}
\end{deluxetable}
	
\end{document}